# *Ragas* in Bollywood music - A microscopic view through multrifractal cross-correlation method

**Shankha Sanyal[a,b]*, Archi Banerjee[a,c], Souparno Roy[a,d], Sayan Nag[e]**

**Ranjan Sengupta[a] and Dipak Ghosh[a]**

*[a]Sir C.V. Raman Centre for Physics and Music, Jadavpur University, Kolkata, India*
*[b] School of Languages and Linguistics, Jadavpur University, Kolkata, India*
*[c]Rekhi Centre of Excellence for the Science of Happiness, IIT Kharagpur, India*
*[d] Department of Physics, Jadavpur University, Kolkata, India*
*[e] Department of Medical Biophysics, University of Toronto, Canada*
**email: archibanerjee7@gmail.com*

## ABSTRACT

*Since the start of Indian cinema, a number of films have been made where a particular song is based on a certain raga. These songs have been taking a major role in spreading the essence of classical music to the common people, who have no formal exposure to classical music. In this paper, we look to explore what are the particular features of a certain raga which make it understandable to common people and enrich the song to a great extent. For this, we chose two common ragas of Hindustani classical music, namely "Bhairav" and "Mian ki Malhar" which are known to have widespread application in popular film music. We have taken 3 minute clips of these two ragas from the renderings of two eminent maestros of Hindustani classical music. 3 min clips of ten (10) widely popular songs of Bollywood films were selected for analysis. These were analyzed with the help of a latest non linear analysis technique called Multifractal Detrended Cross correlation Analysis (MFDXA). With this technique, all parts of the Film music and the renderings from the eminent maestros are analyzed to find out a cross correlation coefficient ($\gamma_x$) which gives the degree of correlation between these two signals. We hypothesize that the parts which have the highest degree of cross correlation are the parts in which that particular raga is established in the song. Also the variation of cross correlation coefficient in the different parts of the two samples gives a measure of the modulation that is executed by the singer. Thus, in nutshell we try to study scientifically the amount of correlation that exists between the raga and the same raga being utilized in Film music. This will help in generating an automated algorithm through which a naïve listener will relish the flavor of a particular raga in a popular film song. The results are discussed in detail.*

## 1. INTRODUCTION

Raga is said to be the soul of Indian music. Hindi film music, which is commonly referred to as "Bollywood" music, is one of the most popular forms of music in the world today. Almost all Bollywood movies feature several songs that are very popular in India, and quite a large section of these songs are based on certain *raga* [1]. The classical music system of Indian sub-continent is based on two major concepts – raga and tala. Raga describes the melodic or modal aspect of music and tala describes the rhythmic aspect. The rhythmic pattern of any composition in Indian music is described by the term tala, which is composed of cycles of matra's. As the major form of mass entertainment available on a national scale, Hindi cinema plays a prominent and influential role in Indian society. Yet its songs, which represent India's most popular music in the twentieth century, have not been subject to any scientific study. Musicians extensively composed a popular song on the basis of a particular Hindustani *raga,* the elements of the *raga* used to be predominant in that song which could be identified by professionals. This paper tries to explore a methodology by which one can understand about the essence of a particular raga applied to the popular Bollywood song. Thus it will be possible also for a novice in raga music, to get a flavor of that raga just by hearing the popular song. This can work as a process of popularization of raga music. For the study, we chose two common ragas namely "Bhairav" and "Mian ki Malhar" of which one is a time based raga conventionally sung in dawn and the other is a seasonal raga depicting monsoon. We have taken 3 minute clips from the '*alaap'* portion of vocal renderings of these two ragas sung by two eminent maestros of Hindustani music. The '*alaap'* portion of the raga rendition is chosen so as to keep the signal free from tempo variation and also normal accompaniment. Moreover, we get the complete raga structure in the *'alaap'* portion. Ten (10) widely popular songs from Bollywood Films are also taken in 3 minute clips [2]. All the music signals were divided into six equal segments of 30 seconds each cut in zero crossing. It is well established that musical signals are generally self-similar. With the possible exception of a few avant-garde compositions, structure and repetition is a general feature of nearly all music [3]. In this context fractal analysis of the signal which reveals the geometry embedded in signal assumes significance.

It is also known that naturally evolving geometries and phenomena are rarely characterized by a single scaling ratio and therefore different parts of a system may be scaling differently. That is, the self similarity pattern is not uniform over the whole system. Such a system is better characterized as 'multifractal'. A multifractal can be loosely thought of as an interwoven set constructed from sub-sets with different local fractal dimensions. Music too, has non-uniform property in its movement. It is therefore necessary to re-investigate the musical structure from the viewpoint of the multifractal theory. Multifractal Detrended Fluctuation Analysis (MFDFA) [4] technique analyzes the musical signal in different scales and gives a multifractal spectral width which is the measure of complexity of the signal. Correlation determines the degree of similarity between two signals. In signal processing, cross-correlation is a measure of similarity of two series as a function of the lag of one relative to the other. A non linear technique called Multifractal Detrended Cross correlation Analysis (MFDXA) [5] is used to analyze the multifractal behaviors in the power-law cross-correlations between two time series data of music signals. With this technique, all segments of the film music and the renderings from the eminent maestros are analyzed to find out a cross correlation coefficient ($\gamma_x$) which gives the degree of correlation between these two categories of signals. For uncorrelated data, $\gamma_x$ has a value 1 and the lower the value of $\gamma_x$ more correlated is the data [6]. Thus a negative value of $\gamma_x$ signifies that the two music signals have very high degree of correlation between them. We hypothesize that the songs which show high degree of cross correlation (i. e lower value of $\gamma_x$) with the raga clips are the songs in which the essence of that particular raga is present. Also the variation of cross correlation coefficient in the different parts of the two samples gives a measure of the modulation that is executed by the singer. Further, the songs having lower degree of cross correlation with the raga clips are the ones in which the characteristics of that particular raga is absent, may be the features of some other raga or a mixture of ragas is present. Thus, in nutshell we propose an automated scientific algorithm with the help of which we can easily check whether the features or the structure of a particular raga is present in popular Hindi film songs.

## 2. EXPERIMENTAL DETAILS

Ten (10) clips of popular Bollywood songs were taken from [2], each of 3 minute duration. The songs chosen for analysis are detailed in Table 1. A 3 min clip from the *alaap* part of the two *ragas, Bhairav* and *Mian ki Malhar* were taken from the rendition of two eminent vocalists of Hindustani music. The signals are digitized at the rate of 22050 samples/sec 16 bit format. The *alaap* part was considered for analysis because the characteristic features of the entire *raga* is present in this part and that it uses all the notes used in that particular raga and allowed transitions between them with proper distribution over time. Each three minutes signal is divided into six equal segments of 30 seconds each. We measure the cross correlation coefficient for each of the six windows and record their variation. The renditions from two different vocalists were chosen to cross check the validation of our hypothesis that the cross correlation coefficient can really be a measure for identification of a particular *raga* on which a song is based on.

| Clip No. | Song Name | Singer | Film (Year) |
|---|---|---|---|
| **Clip 1** | Mohabbat Ki Jhooti Kahani | Lata Mangeshkar | Mughal - E - Azam (1960) |
| **Clip 2** | Karo Sab Nichchawr | Asha Bhosle | Ladki Sayadri Ki (1966) |
| **Clip 3** | Mohe Bhul Gaye Sanvariya | Lata Mangeshkar | Baiju Bawra (1952) |
| **Clip 4** | Naach Mere Mor Zara Naach | Manna Dey | Tere Dwar Khada Bhagwan (1964) |
| **Clip 5** | Dil Diya Dard Liya | Md. Rafi | Dil Diya Dard Liya (1966) |
| **Clip 6** | Na-Na-Na Barso Badal | Lata Mangeshkar | Prithviraj Chauhan (1959) |
| **Clip 7** | Ek Ritu Aaye Ek Ritu | Kishore Kumar | Gautam Govinda (1979) |
| **Clip 8** | Duniya Bananewale | Manna Dey | Ziddi (1964) |
| **Clip 9** | Insaan Bano | Naushad | Baiju Bawra (1952) |
| **Clip 10** | Amma Roti De | Lata Mangeshkar | Sansaar (1952) |

**Table1:** Different Bollywood songs chosen for our analysis

## 3. METHOD OF ANALYSIS

### 3.1. Multifractal Detrended Cross Correlation Analysis (MF-DXA):

We have performed a cross-correlation analysis of correlation between different Bollywood songs and *raga* clips following the prescription of Zhou [5].

$$x_{avg} = 1/N \sum_{i=1}^{N} x(i) \text{ and } y_{avg} = 1/N \sum_{i=1}^{N} y(i) \quad (1)$$

Then we compute the profiles of the underlying data series x(i) and y(i) as

$$X(i) \equiv \left[\sum_{k=1}^{i} x(k) - x_{avg}\right] \text{ for } i = 1 \ldots N \quad (2)$$

$$Y(i) \equiv \left[\sum_{k=1}^{i} x(k) - x_{avg}\right] \text{ for } i = 1 \ldots N \quad (3)$$

The qth order detrended covariance Fq(s) is obtained after averaging over 2Ns bins.

$$F_q(s) = \{1/2N_s \sum_{v=1}^{2Ns} [F(s,v)]^{q/2}\}^{1/q} \quad (4)$$

where q is an index which can take all possible values except zero because in that case the factor 1/q blows up. The procedure can be repeated by varying the value of s. Fq(s) increases with increase in value of s. If the series is long range power correlated, then Fq(s) will show power law behavior

$$F_q(s) \sim s^{\lambda(q)}.$$

Zhou found that for two time series constructed by binomial measure from p-model, there exists the following relationship [5]:

$$\lambda(q = 2) \approx [h_x(q = 2) + h_y(q = 2)]/2. \quad (5)$$

Podobnik and Stanley have studied this relation when q = 2 for monofractal Autoregressive Fractional Moving Average (ARFIMA) signals and EEG time series [7].

In case of two time series generated by using two uncoupled ARFIMA processes, each of both is autocorrelated, but there is no power-law cross correlation with a specific exponent [53]. According to auto-correlation function given by:

$$C(\tau) = \langle [x(i+\tau) - \langle x \rangle][x(i) - \langle x \rangle] \rangle \sim \tau^{-\gamma}. \quad (6)$$

The cross-correlation function can be written as

$$C_x(\tau) = \langle [x(i+\tau) - \langle x \rangle][y(i) - \langle y \rangle] \rangle \sim \tau^{-\gamma_x} \quad (7)$$

where $\gamma$ and $\gamma_x$ are the auto-correlation and cross-correlation exponents, respectively. Due to the non-stationarities and trends superimposed on the collected data direct calculation of these exponents are usually not recommended rather the reliable method to calculate auto-correlation exponent is the DFA method, namely $\gamma = 2 - 2h(q = 2)$ [8]. Recently, Podobnik et al., have demonstrated the relation between cross-correlation exponent, $\gamma_x$ and scaling exponent $\lambda(q)$ derived from $\gamma_x = 2 - 2\lambda(q = 2)$ [7]. For uncorrelated data, $\gamma_x$ has a value 1 and the lower the value of $\gamma$ and $\gamma_x$ more correlated is the data.

In general, $\lambda(q)$ depends on q, indicating the presence of multifractality. In other words, we want to point out how two sound signals are cross-correlated in various time scales i.e. how much is the essence of a particular *raga* present in a specific song.

## 4. RESULTS AND DISCUSSIONS

The cross-correlation coefficient, $\gamma_x$ was computed for each part of the song and the *raga* clips of the two artistes. Lower the value of $\gamma_x$, higher is the degree of cross-correlation between the two signals. A significantly low $\gamma_x$ would thus signify a greater cue of a particular *raga* being present in that song. The human mind identifies the presence of a *raga* in a song by recognizing certain note structures or transitions that is the characteristic of that *raga*. We want to recreate those features in an automated way, whereby a unique parameter will define the presence of a certain *raga* or a mixture of *ragas* in a musical piece. Fig. 1 and 2 represent the variation of $\gamma_x$ for the 10 clips and the rendition of *raga Bhairav* by two artistes while Fig. 3-4 represents the same for *Raga Mia ki Malhar*

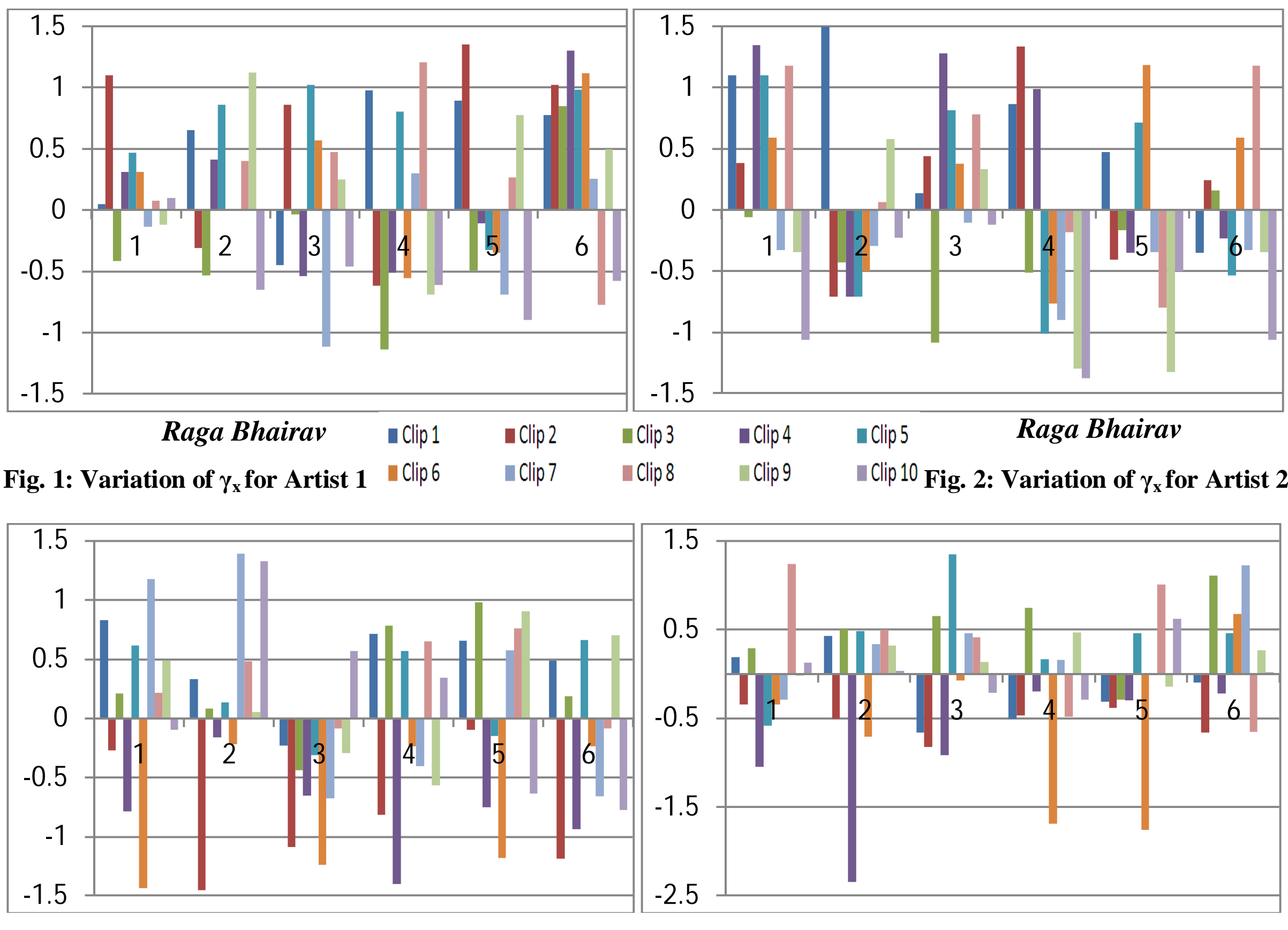

**Fig. 1: Variation of $\gamma_x$ for Artist 1** **Fig. 2: Variation of $\gamma_x$ for Artist 2**

**Fig. 3: Variation of $\gamma_x$ for Artist 1** **Fig. 4: Variation of $\gamma_x$ for Artist 2**

***Raga Mia ki Malhar***

The following observations can be drawn from a careful study of the figures:

1. From Figs. 1 and 2, it is seen that Clips 3, 7 and 10 are the ones which in general have negative values of $\gamma_x$ signifying that they are the ones in which the features of *raga Bhairav* are the strongest. A number of significant dips are also observed in parts of the same song, which can be attributed to very close proximity of the *raga* and the song. We hypothesize that these are the parts in which a person could relate that a song is based on which particular *raga.*
2. Clip 7 in case of Artist 1 shows a certain amount of ambiguity in belonging to *Raga Bhairav* as is seen from Fig. 2, that the value of $\gamma_x$ jumps to positive a number of times. But still, since it has considerable low values in other parts, it was considered as based on *raga Bhairav.*
3. It is seen from Figs. 3 and 4 that Clips 2, 4 and 6 show predominant features of *raga Mia ki Malhar* in them. Part 2 of Clip 4, especially has a noteworthy drop in $\gamma_x$ for Artist 2. This shows that the note structure of *raga Mia ki Malhar* is absolutely followed in this part which results in such a higher degree of cross correlation.
4. Clip 5 shows a unique property of being strong correlated with both the *ragas* in a number of regions. Thus, we can safely consider that there are elements of different *ragas* woven together in a single piece. If the program is repeated for some more *ragas,* we may have glimpses of more *ragas* present in each of the Clips taken for analysis here.
5. It is also seen from the figures that all the other clips, in some part or the other has some negative values of $\gamma_x$, but that may be caused due to spurious matching of note structures of the two clips which has eventually led to higher values of correlation among themselves. It is known that a similar note combination may be used in a number of different *ragas* in Hindustani music, but each *raga* has its own identity which is manifested in the transition between its notes, duration, stress and emphasis on some particular note. All these parameters when taken into consideration will lead to higher degree of cross correlation between two signals which belong to the same *raga.* Hence, we have considered only those clips belonging to a certain *raga* which have consistently shown higher degree of cross-correlation in all the parts.

Next, we have extracted the highest value of cross-correlation from for each combination and put them in **Table 2**, which will give the readers an estimate of which song is based on which *raga.* The maximum value of $\gamma_x$ was computed from the six parts for which the program was carried out.

**Table 2:** Maximum values of $\gamma_x$ for each *raga*

| | **Raga Bhairav** | | **Raga Mia ki Malhar** | |
|---|---|---|---|---|
| | **Artist 1** | **Artist 2** | **Artist 1** | **Artist 2** |
| **Clip 1** | -0.44 | -0.35 | -0.22 | -0.66 |
| **Clip 2** | -0.61 | -0.7 | -1.45 | -0.817 |
| **Clip 3** | -1.13 | -1.08 | -0.44 | -0.28 |
| **Clip 4** | -0.53 | -0.7 | -1.39 | -1.04 |
| **Clip 5** | -0.32 | -0.53 | -0.3 | -0.58 |
| **Clip 6** | -0.34 | -0.56 | -1.43 | -1.75 |
| **Clip 7** | -1.11 | -0.89 | -0.67 | -0.28 |
| **Clip 8** | -0.7 | -0.17 | -0.08 | -0.64 |
| **Clip 9** | -0.11 | -0.34 | -0.56 | -0.14 |
| **Clip 10** | -0.89 | -1.37 | -0.77 | -0.288 |

It is seen from Table 2, that whenever the maximum value of $\gamma_x$ is below -0.8, signature of the presence of a particular *raga* is seen in a musical clip. This is ascertained from our previous observation that these are the clips which showed consistency in maintaining a higher degree of cross correlation in all the parts. This is an interesting observation which can have far reaching consequences when it comes to the study of Hindustani music using non linear analysis. Although, taking the maximum value could be a bit misleading as we are having negative values of $\gamma_x$ for all the Clips, which implies certain amount of correlation is present always. But again, that can be attributed to the use of similar notes/note-note transition in that particular part where we are getting maximum cross-correlation. Nevertheless, using this technique we have succeeded in obtaining a specific baseline value of -0.80, beyond which any value implies the manifestation of a particular *raga* in a song.

## 5. CONCLUSION

According to Naushad Ali, a prominent film music composer, ".. classical music has never been the art of masses. It flourished in glamorous courts of Rajas, Maharajas and Nawabs. The common people, who had no access to the great courts were never offered the opportunity of listening to classical music and therefore could not acquire an ear for appreciating it. Hindi film music gave them ample scope to listen to and appreciate classical music [9]". Thus, it has long been said that popularization of classical music can be done through Bollywood music. This work presents a new, interesting data regarding non-linear multifractal analysis of sound signals and the application in the field of *raga* identification from certain Bollywood songs. The following conclusions can be drawn from the study done:

1. We make use of a robust non linear technique MFDXA to quantify the degree of cross correlation between a *raga* clip and a popular Bollywood song. From the cross correlation coefficient, we get a cue for the presence of a particular *raga* in that song.
2. The cross correlation coefficient also gives us a clue about a number of different *ragas* merged together in a certain song which is difficult to be identified by only auditory perception.
3. We have defined a baseline value, beyond which any song will fall in the category of a particular *raga.* Below the stipulated value, the song will have the flavor of more than one *raga.*

In this age of dwindling popularity of classical music, this study could be a boon for popularizing it in the country. If we could generate a one-to-one relation between the success rate of a certain song and the *raga* present in it, the use of that *raga* in more and more songs could be a welcome solution.

## 6. ACKNOWLEDGEMENTS